%% file: CSquant2.tex
\documentclass[12pt]{article}
\normalsize
\newcommand{\bdi}{\begin{displaymath}}
\newcommand{\edi}{\end{displaymath}}
\newcommand{\bfi}{\begin{figure}}
\newcommand{\efi}{\end{figure}}

\newcommand{\beq}{\begin{equation}}
\newcommand{\eeq}{\end{equation}}
\newcommand{\beqa}{\begin{eqnarray}}
\newcommand{\eeqa}{\end{eqnarray}}

\newcommand{\ra}{\rightarrow}

\newcommand{\Dsla}{D\hspace{-7.3pt}  /  }

\def\longbar#1{\setbox1=\hbox{$#1$}
\setbox2=\vbox{\hrule width 0.8\wd1}
\raise0.5\ht1\hbox{${\lower\dp1\box2}\atop\box1$}}  

\begin{document}

\begin{titlepage}

\begin{flushright}
\today
\end{flushright}

\vspace{1cm}
\begin{center}
{\Large \bf Chern--Simons action for zero-mode supporting gauge fields in 
three dimensions }\\[1cm]
C. Adam* \\
Wolfgang-Pauli Institute c/o Institute of Mathematics, Vienna University,
 1090 Vienna\\

\medskip

\medskip

B. Muratori** \\
CERN, SL-AP Division, Geneva 32, CH-1211

\medskip

\medskip

C. Nash*** \\
Department of Mathematical Physics, National University of Ireland, Maynooth
\vfill
{\bf Abstract} \\
\end{center}

Recent results on zero modes of the Abelian Dirac operator in three dimensions
support to some degree the conjecture that the Chern--Simons action 
admits only certain quantized values for gauge fields that lead to zero
modes of the corresponding Dirac operator. Here we show that this conjecture
is wrong by constructing an explicit counter-example.

\vfill

$^*)${\footnotesize  
email address: adam@mat.univie.ac.at} 

$^{**})${\footnotesize
email address: Bruno.Muratori@cern.ch} 

$^{***})${\footnotesize
email address: cnash@stokes2.thphys.may.ie} 
\end{titlepage}


In the last few years a considerable amount of interest has been devoted to
the study of zero modes of the Abelian Dirac operator in 
three-dimensional Euclidean space, that is, to square-integrable solutions
of the Dirac equation
\beq
 \Dsla \Psi \equiv 
\vec \sigma (i\vec \partial +\vec A (\vec x))\Psi (\vec x) =0
\eeq
where $\vec x =(x_1,x_2,x_3)$, $\vec \sigma$ are the Pauli matrices, and
$\Psi $ is a two-component spinor. In addition, the gauge field $\vec A$
is assumed to obey certain integrability conditions (e.g., square 
integrability of the related magnetic field $\vec B =\vec \partial \times
\vec A$).
 
On the one hand, such solutions are relevant for the quantum mechanical 
behaviour of non-relativistic electrons (see e.g., \cite{FLL}), 
because solutions to the above 
equation are, at the same time, solutions to the Pauli equation (the Pauli
equation is obtained by just squaring the Dirac operator in the above 
equation, i.e., $\Dsla^2 \Psi =0$). 
On the other hand, solutions to the Dirac equation are also relevant for
(Euclidean) quantum electrodynamics, as was discussed, e.g., in 
\cite{Fry1,Fry2}.   

Some first examples of zero modes were constructed in  \cite{LoYa}. In 
\cite{AMN1} a class of Dirac operators and their zero modes was
constructed which depend on a function that is arbitrary up to certain
boundary conditions, thereby relating the existence of these zero modes
to some topological condition. Some further examples of zero modes were given
in \cite{El1} and in \cite{AMN2}. In \cite{AMN3,AMN4} the first examples
of Dirac operators with multiple zero modes were given, thereby demonstrating 
the existence of the phenomenon of zero mode degeneracy. 
Further, a relation  between the
number of zero modes and a certain topological linking number
 (the Hopf index) of the corresponding gauge field was established in
\cite{AMN4}. A very detailed and more geometrical discussion of these
Dirac operators with multiple zero modes, based on the concept of
Riemannian submersions, was given in \cite{ErSo}.

In \cite{BaEv} the following two results were proved: {\em i)} For the
one-parameter family of gauge potentials $t\vec A$ zero modes may exist
for at most a finite set of values $t_i$ for any $t \in (t_0 ,t_1)$, and
{\em ii)} The set of gauge potentials with {\em no} zero modes is a dense
subset of the set of all gauge potentials (with certain decay properties).
Recently, some results on the dimensionality of the space of gauge potential
with zero modes were obtained in \cite{El2}. There it was proven that locally
the space of gauge potentials with (at least) one zero mode is of co-dimension
one within the space of all gauge potentials (with certain decay 
properties). In addition, some results on
the dimensionalities of spaces of gauge potentials with multiple zero modes
were proven. 

The above-described results would suit well with the assumption that there
exists a certain functional of the gauge potential which may admit only
fixed or quantized values for gauge potentials that support zero modes.
The simplest functional one can imagine is the Chern--Simons action, 
which has the additional attractive feature of being a topological invariant
(i.e., independent of the metric). Therefore, if the existence and 
degeneracy of zero
modes is related to some topological features, as was speculated, e.g., in
\cite{Fry2}, the Chern--Simons action would be an obvious candidate.  

In addition, the assumption of
quantized Chern--Simons action for gauge potentials with zero modes is
further supported by the results of \cite{AMN4}, where a whole class
of gauge potentials with an arbitrary number of zero modes was constructed.
For all these gauge potentials, which are characterized by an arbitrary 
function and an integer $l$ (the number of zero modes for a given gauge 
potential), the Chern--Simons action indeed admits only the quantized 
values
\beq \label{CS-qu}
\frac{1}{16 \pi^2}\int d^3 x \vec A\cdot \vec B =\frac{1}{4}(l+\frac{1}{2}
)^2
\eeq
where $l$ is the number of zero modes for the given gauge potential
(the Chern--Simons action for all the gauge potentials of \cite{AMN4}
was calculated explicitly in \cite{AMN5}).

[Remark: In \cite{AMN4} a fixed universal background gauge potential was
added to all the gauge potentials, $\vec A \ra \vec {\tilde A} = \vec A +
\vec A^{\rm b}$, in order to relate the resulting gauge potentials 
$ \vec {\tilde A}$ to Hopf maps. For these resulting gauge potentials the
Chern--Simons action is automatically quantized, 
$ (1/16 \pi^2)\int d^3 x \vec {\tilde A}\cdot 
\vec {\tilde B} =(1/4)(l+1)^2$, because the
integer Hopf index is given by the Chern--Simons action. However, even the
original zero-mode supporting gauge potentials, without the background
field, lead to the quantized Chern--Simons action (\ref{CS-qu}), although
they cannot be directly related to Hopf maps.]  

Another, more general argument in favour of quantized Chern--Simons action
for zero-mode supporting gauge fields is related to the anomaly equation
in Minkowski space
\beq
\partial^\mu J_\mu =\frac{1}{16 \pi^2}
\epsilon^{\mu\nu\alpha\beta}F_{\mu\nu}F_{\alpha\beta} \, .
\eeq
Here, $J_\mu$ is the current density of a chiral (left-handed) Weyl
Fermion which couples to the Abelian gauge field $A_\mu$ with field
strength $F_{\mu\nu}$. Choosing the gauge $A_0=0$ and integrating the
above equation over all space and over the finite time interval $[t_i,
t_f]$ results in the equation
\beq \label{part-no}
Q(t_f) -Q (t_i) =-\frac{1}{4\pi} (I_{\rm CS}(t_f) - I_{\rm CS}(t_i))
\eeq
\beq
Q(t)\equiv \int d^3 x J_0 (t,\vec x) \quad ,\quad I_{\rm CS} (t) \equiv 
\int d^3 x (\vec A\cdot \vec B)(t_,\vec x)
\eeq
where the l.h.s. of (\ref{part-no})
is the change in particle number between $t_{\rm i}$
and $t_{\rm f}$ \cite{AMN5}.
If we further assume that the change in particle number, $Q(t_f) -Q (t_i)$,
is two times the number of levels (zero modes) $L$ crossed in the (adiabatic)
change from $\vec A(t_i ,\vec x)$ to $\vec A (t_f,\vec x)$ - as is usually
assumed - then the equation
\beq \label{level}
2L  =-\frac{1}{4\pi} (I_{\rm CS}(t_f) - I_{\rm CS}(t_i))
\eeq 
results. Consequently, $L$ must be independent of the path 
$\vec A(t,\vec x)$ which connects
$\vec A(t_i,\vec x)$ and $\vec A(t_f,\vec x)$ 
(see \cite{AMN5} and the literature cited there for details). 
Obviously, Eq. (\ref{level}) is automatically satisfied when level crossing
may occur only for fixed, quantized values of the Chern--Simons action
$I_{\rm CS}$. Therefore, Eq. (\ref{level}) is compatible with the
assumption that zero modes only exist for certain quantized values of
the Chern--Simons action. The correctness of this assumption would, in fact,
be the simplest way to realize Eq. (\ref{level}). 

All in all, the existing results on zero modes provide some evidence
for the assumption that the Chern--Simons action is quantized
for gauge potentials with zero modes. 
Therefore, a further investigation of this question is of some interest.
Topologically the Chern--Simons action 
corresponds to a cohomology class in $H^3(S^3;{\bf R/Z})$ and this is not 
constrained to be discrete. In the remainder of the paper 
we want to demonstrate that the above
 assumption is incorrect in the general case by constructing a simple
counter-example. 

For this purpose, let us briefly review some results from \cite{AMN1}.
There it was shown that the ansatz ($r\equiv |\vec x|$)
\beq
\Psi =g(r)\exp (if(r)\frac{\vec x}{r} \vec\sigma )
\left( \begin{array}{c} 1  
\\ 0 \end{array} \right)
\eeq
for the spinor leads to a zero mode for the gauge field
\beq
A_i =h(r) \frac{\Psi^\dagger \sigma_i \Psi}{\Psi^\dagger \Psi} 
\eeq
provided that $g(r)$ and $h(r)$ are given in terms of the independent
function $f(r)$ as ($' \equiv d/dr$)
 \beq
g' =-\frac{2}{r}\frac{t^2}{1+t^2}g.
\eeq
\beq
h=(1+t^2)^{-1}(t' +\frac{2}{r}t)
\eeq
where
\beq
t(r):= \tan f(r).
\eeq
Here, a sufficient condition on $t(r)$ leading to
smooth, non-singular and $L^2$ spinors and
smooth, non-singular gauge potentials with finite energy ($\int (\vec
B)^2$) and finite Chern--Simons action ($\int \vec A \vec B$) is
\beq \label{int-nul}
t(0)=0 \, , \quad t(r) \sim c_1 r + O(r^2) \quad  {\rm for} \quad r \to 0
\eeq
\beq \label{int-inf}
t(\infty) =\infty .
\eeq
Further, the Chern--Simons action for this class of gauge fields may be
expressed like
\bdi
\int d^3 x \vec A\cdot \vec B = 
4\pi (2\pi - 4) \int_0^\infty dr
\frac{rth^2}{1+t^2} + 4\pi (-2\pi +8) \int_0^\infty dr
\frac{r^2 t' h^2}{1+t^2} =
\edi
\beq
4\pi \int_0^\infty \frac{dr}{r}(rf' +\sin 2f )^2 
[(-2\pi +8) rf' +(\pi -2)\sin 2f ]
\eeq
where spherical coordinates $(r,\theta ,\varphi )$ were introduced and the
angular integrations have already been performed. 

Next, let us introduce the one-parameter family of functions
\beq
t_a (r) =r (a+r^2) 
\eeq
where $a$ is an arbitrary real number. $t_a$ obeys the integrability 
conditions (\ref{int-nul}) and (\ref{int-inf}) for all values of $a$, 
therefore $t_a$ defines a zero-mode supporting gauge potential and a
well-behaving zero mode for all values of $a$. 
We easily compute
\bdi
t_a' = a+3r^2 \quad , \quad h_a = \frac{3a+5r^2}{1+r^2 (a+r^2)^2}
\edi
and
\beq \label{CS-ac-a}
\int d^3 x \vec A\cdot \vec B = 16\pi \int_0^\infty dr r^2 [a+(5-\pi)r^2]
\frac{(3a+5r^2)^2}{[1+r^2 (a+r^2)^2]^3} .
\eeq  
This last integral may be easily evaluated numerically with the help of
Mathematica. Before presenting the result of the numerical integration, 
we want to remark that, in the limit 
$a\to +\infty$, $t_a$ is equal to the simplest case $t (r) =r$ up to an
(infinite) rescaling, which does not change the value of the Chern--Simons 
action. Further, the simplest gauge potential with $t=r$ is, 
at the same time,
the simplest gauge potential of the class of gauge potentials
constructed in \cite{AMN4}, with
one zero mode (i.e., $l=1$). The resulting Chern--Simons action may be
evaluated with the help of (\ref{CS-qu}), and leads to 
\beq
\int d^3 x \vec A\cdot \vec B = 16\pi^2 \frac{1}{4}\left( \frac{3}{2}
\right)^2 = 9\pi^2 .
\eeq
Therefore, the Chern--Simons action (\ref{CS-ac-a})
normalized according to $(9\pi^2)^{-1}\int
d^3 x \vec A\cdot \vec B$ should approach the value one in the limit of large
positive $a$. 

In Fig. 1 we plot the Chern--Simons action (\ref{CS-ac-a}) multiplied
by $(9\pi^2)^{-1}$, as a function of $a$. 
For $a\to +\infty$ it indeed approaches the value one.
Further, it increases for decreasing values of $a$ and reaches arbitrarily 
large values for sufficiently negative values of $a$. Obviously, the
Chern--Simons action in Fig. 1 may admit all values greater than one.
This result demonstrates that the Chern--Simons action for gauge potentials
with zero modes is, in general, {\em not} quantized, which was the purpose
of this brief report.

\bigskip

\hspace*{-0.7cm} {\large\bf Acknowledgment:} 
The authors thank R. Jackiw for helpful comments. Further,
CA acknowledges support from the Austrian 
START award project FWF-Y-137-TEC of N.J. Mauser.

\newpage
\input psbox.tex
\begin{figure}
$$ \psboxscaled{1100}{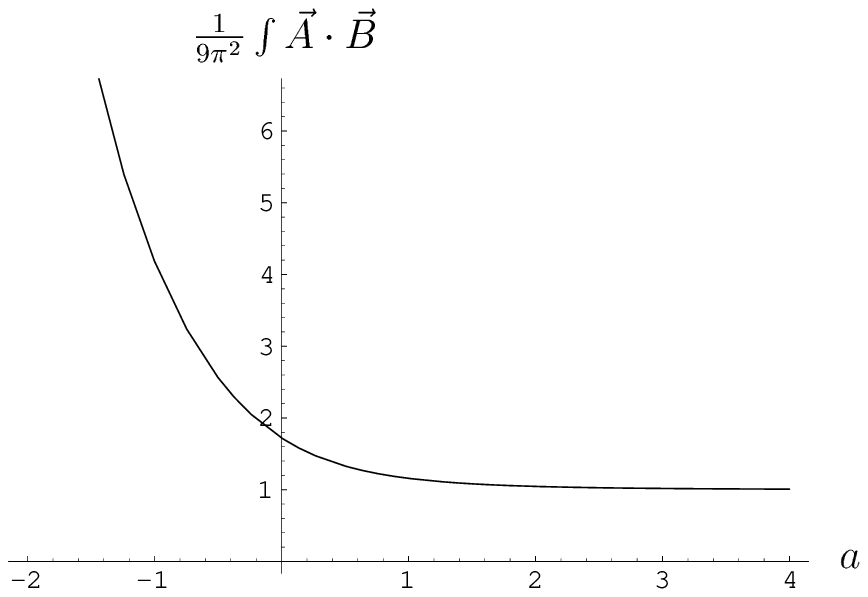} $$
\caption{The appropriately normalized Chern--Simons action $\frac{1}{9\pi^2}
\int \vec A\cdot \vec B$ as a function of the parameter $a$.
}
\end{figure}

\end{document}

%% file: psbox.tex
\def\temp{1.34}%
\let\tempp=\relax
\expandafter\ifx\csname psboxversion\endcsname\relax
  \message{PSBOX(\temp) loading}%
\else
    \ifdim\temp cm>\psboxversion cm
      \message{PSBOX(\temp) loading}%
    \else
      \message{PSBOX(\psboxversion) is already loaded: I won't load
        PSBOX(\temp)!}%
      \let\temp=\psboxversion
      \let\tempp= 
    \fi
\fi
\tempp
\let\psboxversion=\temp
\catcode`\@=11
%
%
\def\psfortextures{
\def\PSspeci@l##1##2{%
\special{illustration ##1\space scaled ##2}%
}}%
\def\psfordvitops{
\def\PSspeci@l##1##2{%
\special{dvitops: import ##1\space \the\drawingwd \the\drawinght}%
}}%
\def\psfordvips{
\def\PSspeci@l##1##2{%
\d@my=0.1bp \d@mx=\drawingwd \divide\d@mx by\d@my
\includegraphics{##1\space}}}%
\def\psforoztex{
\def\PSspeci@l##1##2{%
\special{##1 \space
      ##2 1000 div dup scale
      \number-\psllx\space \number-\pslly\space translate
}}}%
\def\psfordvitps{
\def\psdimt@n@sp##1{\d@mx=##1\relax\edef\psn@sp{\number\d@mx}}
\def\PSspeci@l##1##2{%
\special{dvitps: Include0 "psfig.psr"}
\psdimt@n@sp{\drawingwd}
\special{dvitps: Literal "\psn@sp\space"}
\psdimt@n@sp{\drawinght}
\special{dvitps: Literal "\psn@sp\space"}
\psdimt@n@sp{\psllx bp}
\special{dvitps: Literal "\psn@sp\space"}
\psdimt@n@sp{\pslly bp}
\special{dvitps: Literal "\psn@sp\space"}
\psdimt@n@sp{\psurx bp}
\special{dvitps: Literal "\psn@sp\space"}
\psdimt@n@sp{\psury bp}
\special{dvitps: Literal "\psn@sp\space startTexFig\space"}
\special{dvitps: Include1 "##1"}
\special{dvitps: Literal "endTexFig\space"}
}}%
\def\psfordvialw{
\def\PSspeci@l##1##2{
\special{language "PostScript",
position = "bottom left",
literal "  \psllx\space \pslly\space translate
  ##2 1000 div dup scale
  -\psllx\space -\pslly\space translate",
include "##1"}
}}%
\def\psforptips{
\def\PSspeci@l##1##2{{
\d@mx=\psurx bp
\advance \d@mx by -\psllx bp
\divide \d@mx by 1000\multiply\d@mx by \xscale
\incm{\d@mx}
\let\tmpx\dimincm
\d@my=\psury bp
\advance \d@my by -\pslly bp
\divide \d@my by 1000\multiply\d@my by \xscale
\incm{\d@my}
\let\tmpy\dimincm
\d@mx=-\psllx bp
\divide \d@mx by 1000\multiply\d@mx by \xscale
\d@my=-\pslly bp
\divide \d@my by 1000\multiply\d@my by \xscale
\at(\d@mx;\d@my){\special{ps:##1 x=\tmpx, y=\tmpy}}
}}}%
\def\psonlyboxes{
\def\PSspeci@l##1##2{%
\at(0cm;0cm){\boxit{\vbox to\drawinght
  {\vss\hbox to\drawingwd{\at(0cm;0cm){\hbox{({\tt##1})}}\hss}}}}
}}%
\def\psloc@lerr#1{%
\let\savedPSspeci@l=\PSspeci@l%
\def\PSspeci@l##1##2{%
\at(0cm;0cm){\boxit{\vbox to\drawinght
  {\vss\hbox to\drawingwd{\at(0cm;0cm){\hbox{({\tt##1}) #1}}\hss}}}}
\let\PSspeci@l=\savedPSspeci@l
}}%
%
%
\newread\pst@mpin
\newdimen\drawinght\newdimen\drawingwd
\newdimen\psxoffset\newdimen\psyoffset
\newbox\drawingBox
\newcount\xscale \newcount\yscale \newdimen\pscm\pscm=1cm
\newdimen\d@mx \newdimen\d@my
\newdimen\pswdincr \newdimen\pshtincr
\let\ps@nnotation=\relax
{\catcode`\|=0 |catcode`|\=12 |catcode`|
|catcode`#=12 |catcode`*=14
|xdef|backslashother{\}*
|xdef|percentother{
|xdef|tildeother{~}*
|xdef|sharpother{#}*
}%
\def\R@moveMeaningHeader#1:->{}%
\def\uncatcode#1{%
\edef#1{\expandafter\R@moveMeaningHeader\meaning#1}}%
\def\execute#1{#1}
\def\psm@keother#1{\catcode`#112\relax}
\def\executeinspecs#1{%
\execute{\begingroup\let\do\psm@keother\dospecials\catcode`\^^M=9#1\endgroup}}%
\def\@mpty{}%
\def\matchexpin#1#2{
  \fi%
  \edef\tmpb{{#2}}%
  \expandafter\makem@tchtmp\tmpb%
  \edef\tmpa{#1}\edef\tmpb{#2}%
  \expandafter\expandafter\expandafter\m@tchtmp\expandafter\tmpa\tmpb\endm@tch%
  \if\match%
}%
\def\matchin#1#2{%
  \fi%
  \makem@tchtmp{#2}%
  \m@tchtmp#1#2\endm@tch%
  \if\match%
}%
\def\makem@tchtmp#1{\def\m@tchtmp##1#1##2\endm@tch{%
  \def\tmpa{##1}\def\tmpb{##2}\let\m@tchtmp=\relax%
  \ifx\tmpb\@mpty\def\match{YN}%
  \else\def\match{YY}\fi%
}}%
\def\incm#1{{\psxoffset=1cm\d@my=#1
 \d@mx=\d@my
  \divide\d@mx by \psxoffset
  \xdef\dimincm{\number\d@mx.}
  \advance\d@my by -\number\d@mx cm
  \multiply\d@my by 100
 \d@mx=\d@my
  \divide\d@mx by \psxoffset
  \edef\dimincm{\dimincm\number\d@mx}
  \advance\d@my by -\number\d@mx cm
  \multiply\d@my by 100
 \d@mx=\d@my
  \divide\d@mx by \psxoffset
  \xdef\dimincm{\dimincm\number\d@mx}
}}%
%
\newif\ifNotB@undingBox
\newhelp\PShelp{Proceed: you'll have a 5cm square blank box instead of
your graphics (Jean Orloff).}%
\def\s@tsize#1 #2 #3 #4\@ndsize{
  \def\psllx{#1}\def\pslly{#2}%
  \def\psurx{#3}\def\psury{#4}
  \ifx\psurx\@mpty\NotB@undingBoxtrue
  \else
    \drawinght=#4bp\advance\drawinght by-#2bp
    \drawingwd=#3bp\advance\drawingwd by-#1bp
  \fi
  }%
\def\sc@nBBline#1:#2\@ndBBline{\edef\p@rameter{#1}\edef\v@lue{#2}}%
\def\g@bblefirstblank#1#2:{\ifx#1 \else#1\fi#2}%
{\catcode`\%=12
\xdef\B@undingBox{
\def\ReadPSize#1{
 \readfilename#1\relax
 \let\PSfilename=\lastreadfilename
 \openin\pst@mpin=#1\relax
 \ifeof\pst@mpin \errhelp=\PShelp
   \errmessage{I haven't found your postscript file (\PSfilename)}%
   \psloc@lerr{was not found}%
   \s@tsize 0 0 142 142\@ndsize
   \closein\pst@mpin
 \else
   \if\matchexpin{\GlobalInputList}{, \lastreadfilename}%
   \else\xdef\GlobalInputList{\GlobalInputList, \lastreadfilename}%
     \immediate\write\psbj@inaux{\lastreadfilename,}%
   \fi%
   \loop
     \executeinspecs{\catcode`\ =10\global\read\pst@mpin to\n@xtline}%
     \ifeof\pst@mpin
       \errhelp=\PShelp
       \errmessage{(\PSfilename) is not an Encapsulated PostScript File:
           I could not find any \B@undingBox: line.}%
       \edef\v@lue{0 0 142 142:}%
       \psloc@lerr{is not an EPSFile}%
       \NotB@undingBoxfalse
     \else
       \expandafter\sc@nBBline\n@xtline:\@ndBBline
       \ifx\p@rameter\B@undingBox\NotB@undingBoxfalse
         \edef\t@mp{%
           \expandafter\g@bblefirstblank\v@lue\space\space\space}%
         \expandafter\s@tsize\t@mp\@ndsize
       \else\NotB@undingBoxtrue
       \fi
     \fi
   \ifNotB@undingBox\repeat
   \closein\pst@mpin
 \fi
\message{#1}%
}%
%
%
\def\psboxto(#1;#2)#3{\vbox{%
   \ReadPSize{#3}%
   \advance\pswdincr by \drawingwd
   \advance\pshtincr by \drawinght
   \divide\pswdincr by 1000
   \divide\pshtincr by 1000
   \d@mx=#1
   \ifdim\d@mx=0pt\xscale=1000
         \else \xscale=\d@mx \divide \xscale by \pswdincr\fi
   \d@my=#2
   \ifdim\d@my=0pt\yscale=1000
         \else \yscale=\d@my \divide \yscale by \pshtincr\fi
   \ifnum\yscale=1000
         \else\ifnum\xscale=1000\xscale=\yscale
                    \else\ifnum\yscale<\xscale\xscale=\yscale\fi
              \fi
   \fi
   \divide\drawingwd by1000 \multiply\drawingwd by\xscale
   \divide\drawinght by1000 \multiply\drawinght by\xscale
   \divide\psxoffset by1000 \multiply\psxoffset by\xscale
   \divide\psyoffset by1000 \multiply\psyoffset by\xscale
   \global\divide\pscm by 1000
   \global\multiply\pscm by\xscale
   \multiply\pswdincr by\xscale \multiply\pshtincr by\xscale
   \ifdim\d@mx=0pt\d@mx=\pswdincr\fi
   \ifdim\d@my=0pt\d@my=\pshtincr\fi
   \message{scaled \the\xscale}%
 \hbox to\d@mx{\hss\vbox to\d@my{\vss
   \global\setbox\drawingBox=\hbox to 0pt{\kern\psxoffset\vbox to 0pt{%
      \kern-\psyoffset
      \PSspeci@l{\PSfilename}{\the\xscale}%
      \vss}\hss\ps@nnotation}%
   \global\wd\drawingBox=\the\pswdincr
   \global\ht\drawingBox=\the\pshtincr
   \global\drawingwd=\pswdincr
   \global\drawinght=\pshtincr
   \baselineskip=0pt
   \copy\drawingBox
 \vss}\hss}%
  \global\psxoffset=0pt
  \global\psyoffset=0pt
  \global\pswdincr=0pt
  \global\pshtincr=0pt 
  \global\pscm=1cm 
}}%
%
%
\def\psboxscaled#1#2{\vbox{%
  \ReadPSize{#2}%
  \xscale=#1
  \message{scaled \the\xscale}%
  \divide\pswdincr by 1000 \multiply\pswdincr by \xscale
  \divide\pshtincr by 1000 \multiply\pshtincr by \xscale
  \divide\psxoffset by1000 \multiply\psxoffset by\xscale
  \divide\psyoffset by1000 \multiply\psyoffset by\xscale
  \divide\drawingwd by1000 \multiply\drawingwd by\xscale
  \divide\drawinght by1000 \multiply\drawinght by\xscale
  \global\divide\pscm by 1000
  \global\multiply\pscm by\xscale
  \global\setbox\drawingBox=\hbox to 0pt{\kern\psxoffset\vbox to 0pt{%
     \kern-\psyoffset
     \PSspeci@l{\PSfilename}{\the\xscale}%
     \vss}\hss\ps@nnotation}%
  \advance\pswdincr by \drawingwd
  \advance\pshtincr by \drawinght
  \global\wd\drawingBox=\the\pswdincr
  \global\ht\drawingBox=\the\pshtincr
  \global\drawingwd=\pswdincr
  \global\drawinght=\pshtincr
  \baselineskip=0pt
  \copy\drawingBox
  \global\psxoffset=0pt
  \global\psyoffset=0pt
  \global\pswdincr=0pt
  \global\pshtincr=0pt 
  \global\pscm=1cm
}}%
%
\def\psbox#1{\psboxscaled{1000}{#1}}%
\newif\ifn@teof\n@teoftrue
\newif\ifc@ntrolline
\newif\ifmatch
\newread\j@insplitin
\newwrite\j@insplitout
\newwrite\psbj@inaux
\immediate\openout\psbj@inaux=psbjoin.aux
\immediate\write\psbj@inaux{\string\joinfiles}%
\immediate\write\psbj@inaux{\jobname,}%
%
%
\def\toother#1{\ifcat\relax#1\else\expandafter%
  \toother@ux\meaning#1\endtoother@ux\fi}%
\def\toother@ux#1 #2#3\endtoother@ux{\def\tmp{#3}%
  \ifx\tmp\@mpty\def\tmp{#2}\let\next=\relax%
  \else\def\next{\toother@ux#2#3\endtoother@ux}\fi%
\next}%
%
%
\let\readfilenamehook=\relax
\def\re@d{\expandafter\re@daux}
\def\re@daux{\futurelet\nextchar\stopre@dtest}%
\def\re@dnext{\xdef\lastreadfilename{\lastreadfilename\nextchar}%
  \afterassignment\re@d\let\nextchar}%
\def\stopre@d{\egroup\readfilenamehook}%
\def\stopre@dtest{%
  \ifcat\nextchar\relax\let\nextread\stopre@d
  \else
    \ifcat\nextchar\space\def\nextread{%
      \afterassignment\stopre@d\chardef\nextchar=`}%
    \else\let\nextread=\re@dnext
      \toother\nextchar
      \edef\nextchar{\tmp}%
    \fi
  \fi\nextread}%
\def\readfilename{\bgroup%
  \let\\=\backslashother \let\%=\percentother \let\~=\tildeother
  \let\#=\sharpother \xdef\lastreadfilename{}%
  \re@d}%
%
%
\xdef\GlobalInputList{\jobname}%
\def\psnewinput{%
  \def\readfilenamehook{
    \if\matchexpin{\GlobalInputList}{, \lastreadfilename}%
    \else\xdef\GlobalInputList{\GlobalInputList, \lastreadfilename}%
      \immediate\write\psbj@inaux{\lastreadfilename,}%
    \fi%
    \ps@ldinput\lastreadfilename\relax%
    \let\readfilenamehook=\relax%
  }\readfilename%
}%
\expandafter\ifx\csname @@input\endcsname\relax    
  \immediate\let\ps@ldinput=\input\def\input{\psnewinput}%
\else
  \immediate\let\ps@ldinput=\@@input
  \def\@@input{\psnewinput}%
\fi%
\def\nowarnopenout{%
 \def\warnopenout##1##2{%
   \readfilename##2\relax
   \message{\lastreadfilename}%
   \immediate\openout##1=\lastreadfilename\relax}}%
\def\warnopenout#1#2{%
 \readfilename#2\relax
 \def\t@mp{TrashMe,psbjoin.aux,psbjoint.tex,}\uncatcode\t@mp
 \if\matchexpin{\t@mp}{\lastreadfilename,}%
 \else
   \immediate\openin\pst@mpin=\lastreadfilename\relax
   \ifeof\pst@mpin
     \else
     \errhelp{If the content of this file is so precious to you, abort (ie
press x or e) and rename it before retrying.}%
     \errmessage{I'm just about to replace your file named \lastreadfilename}%
   \fi
   \immediate\closein\pst@mpin
 \fi
 \message{\lastreadfilename}%
 \immediate\openout#1=\lastreadfilename\relax}%
{\catcode`\%=12\catcode`\*=14
\gdef\splitfile#1{*
 \readfilename#1\relax
 \immediate\openin\j@insplitin=\lastreadfilename\relax
 \ifeof\j@insplitin
   \message{! I couldn't find and split \lastreadfilename!}*
 \else
   \immediate\openout\j@insplitout=TrashMe
   \message{< Splitting \lastreadfilename\space into}*
   \loop
     \ifeof\j@insplitin
       \immediate\closein\j@insplitin\n@teoffalse
     \else
       \n@teoftrue
       \executeinspecs{\global\read\j@insplitin to\spl@tinline\expandafter
         \ch@ckbeginnewfile\spl@tinline
       \ifc@ntrolline
       \else
         \toks0=\expandafter{\spl@tinline}*
         \immediate\write\j@insplitout{\the\toks0}*
       \fi
     \fi
   \ifn@teof\repeat
   \immediate\closeout\j@insplitout
 \fi\message{>}*
}*
\gdef\ch@ckbeginnewfile#1
 \def\t@mp{#1}*
 \ifx\@mpty\t@mp
   \def\t@mp{#3}*
   \ifx\@mpty\t@mp
     \global\c@ntrollinefalse
   \else
     \immediate\closeout\j@insplitout
     \warnopenout\j@insplitout{#2}*
     \global\c@ntrollinetrue
   \fi
 \else
   \global\c@ntrollinefalse
 \fi}*
\gdef\joinfiles#1\into#2{*
 \message{< Joining following files into}*
 \warnopenout\j@insplitout{#2}*
 \message{:}*
 {*
 \edef\w@##1{\immediate\write\j@insplitout{##1}}*
\w@{
\w@{
\w@{
\w@{
\w@{
\w@{
\w@{
\w@{
\w@{
\w@{
\w@{\string\input\space psbox.tex}*
\w@{\string\splitfile{\string\jobname}}*
\w@{\string\let\string\autojoin=\string\relax}*
}*
 \expandafter\tre@tfilelist#1, \endtre@t
 \immediate\closeout\j@insplitout
 \message{>}*
}*
\gdef\tre@tfilelist#1, #2\endtre@t{*
 \readfilename#1\relax
 \ifx\@mpty\lastreadfilename
 \else
   \immediate\openin\j@insplitin=\lastreadfilename\relax
   \ifeof\j@insplitin
     \errmessage{I couldn't find file \lastreadfilename}*
   \else
     \message{\lastreadfilename}*
     \immediate\write\j@insplitout{
     \executeinspecs{\global\read\j@insplitin to\oldj@ininline}*
     \loop
       \ifeof\j@insplitin\immediate\closein\j@insplitin\n@teoffalse
       \else\n@teoftrue
         \executeinspecs{\global\read\j@insplitin to\j@ininline}*
         \toks0=\expandafter{\oldj@ininline}*
         \let\oldj@ininline=\j@ininline
         \immediate\write\j@insplitout{\the\toks0}*
       \fi
     \ifn@teof
     \repeat
   \immediate\closein\j@insplitin
   \fi
   \tre@tfilelist#2, \endtre@t
 \fi}*
}%
\def\autojoin{%
 \immediate\write\psbj@inaux{\string\into{psbjoint.tex}}%
 \immediate\closeout\psbj@inaux
 \expandafter\joinfiles\GlobalInputList\into{psbjoint.tex}%
}%
%
%
%
\def\centinsert#1{\midinsert\line{\hss#1\hss}\endinsert}%
\def\psannotate#1#2{\vbox{%
  \def\ps@nnotation{#2\global\let\ps@nnotation=\relax}#1}}%
\def\pscaption#1#2{\vbox{%
   \setbox\drawingBox=#1
   \copy\drawingBox
   \vskip\baselineskip
   \vbox{\hsize=\wd\drawingBox\setbox0=\hbox{#2}%
     \ifdim\wd0>\hsize
       \noindent\unhbox0\tolerance=5000
    \else\centerline{\box0}%
    \fi
}}}%
%
\def\at(#1;#2)#3{\setbox0=\hbox{#3}\ht0=0pt\dp0=0pt
  \rlap{\kern#1\vbox to0pt{\kern-#2\box0\vss}}}%
%
\newdimen\gridht \newdimen\gridwd
\def\gridfill(#1;#2){%
  \setbox0=\hbox to 1\pscm
  {\vrule height1\pscm width.4pt\leaders\hrule\hfill}%
  \gridht=#1
  \divide\gridht by \ht0
  \multiply\gridht by \ht0
  \gridwd=#2
  \divide\gridwd by \wd0
  \multiply\gridwd by \wd0
  \advance \gridwd by \wd0
  \vbox to \gridht{\leaders\hbox to\gridwd{\leaders\box0\hfill}\vfill}}%
%
\def\fillinggrid{\at(0cm;0cm){\vbox{%
  \gridfill(\drawinght;\drawingwd)}}}%
%
%
\def\textleftof#1:{%
  \setbox1=#1
  \setbox0=\vbox\bgroup
    \advance\hsize by -\wd1 \advance\hsize by -2em}%
\def\textrightof#1:{%
  \setbox0=#1
  \setbox1=\vbox\bgroup
    \advance\hsize by -\wd0 \advance\hsize by -2em}%
\def\endtext{%
  \egroup
  \hbox to \hsize{\valign{\vfil##\vfil\cr%
\box0\cr%
\noalign{\hss}\box1\cr}}}%
%
\def\frameit#1#2#3{\hbox{\vrule width#1\vbox{%
  \hrule height#1\vskip#2\hbox{\hskip#2\vbox{#3}\hskip#2}%
        \vskip#2\hrule height#1}\vrule width#1}}%
\def\boxit#1{\frameit{0.4pt}{0pt}{#1}}%
\catcode`\@=12 
%
 \psfordvips   

%% file: CSquant2.bbl
\begin{thebibliography}{99}
\bibitem{FLL}
J. Fr\"ohlich, E. Lieb, M. Loss, 
``Stability of Coulomb systems with magnetic fields 
I'', Commun. Math. Phys. 104 (1986) 251.
\bibitem{Fry1}
M. P. Fry, ``QED in inhomogeneous magnetic fields'',
Phys. Rev. D54 (1996) 6444. 
\bibitem{Fry2}
M. P. Fry, ``Paramagnetism, zero modes and mass singularities in QED
in 1+1 dimensions, 2+1 dimensions and 3+1 dimensions'',
Phys. Rev. D55 (1997) 968; Erratum ibid. D56 (1997) 6714. 
\bibitem{LoYa}
M. Loss, H.-T. Yau, ``Stability of Coulomb systems with magnetic fields 
III. Zero energy bound states of the Pauli operator'', 
Commun. Math. Phys. 104 (1986) 283.
\bibitem{AMN1}
C. Adam, B. Muratori, C. Nash, ``Zero modes of the Dirac operator
in three dimensions'', 
Phys. Rev. D60 (1999) 125001.
\bibitem{El1}
D. M. Elton, ``New examples of zero modes'',
      J. Phys. A33 (41) (2000) 7297.
\bibitem{AMN2}
C. Adam, B. Muratori, C. Nash, ``Zero modes in finite range
  magnetic fields'', Mod. Phys. Lett. A15 (2000) 1577.
\bibitem{AMN3}
C. Adam, B. Muratori, C. Nash, ``Degeneracy of 
zero modes of the Dirac operator in three dimensions'', 
 Phys. Lett. B485 (2000) 314.
\bibitem{AMN4}
C. Adam, B. Muratori, C. Nash, ``Multiple
zero modes of the Dirac operator in three dimensions'',
Phys. Rev. D62 (2000) 085026.
\bibitem{ErSo}
L. Erdos, J. P. Solovej,
``The kernel of Dirac operators on S-3 and R-3'',
Rev. Math. Phys. 13 (10) (2001) 1247. 
\bibitem{BaEv}
A. A. Balinsky, W. D.  Evans, ``On the zero modes of Pauli operators'',
J. Funct. Anal. 179 (1) (2001) 120.
\bibitem{El2}
D. M. Elton, `` The local structure of zero mode producing magnetic 
potentials'', Commun. Math. Phys. 229 (1) (2002) 121.
\bibitem{AMN5}
C. Adam, B. Muratori, C. Nash, ``Particle creation via relaxing
  hypermagnetic knots'', Phys. Rev. D62 (2000) 105027.


\end{thebibliography}
